\documentstyle{article}
\title{Turbulent helium gas cell as a new paradigm of
daily meteorological fluctuations?}
\author{I.M. J\'anosi$^1$, G. Vattay$^{2}$ and A. 
Harnos$^2$}
\begin{document}
\maketitle
\begin{flushleft}
$^1$Physics of Complex Systems Group, E\"otv\"os University, H-1088 Budapest, M\'uzeum krt. 6-8, Hungary\\
$^2$University of Veterinary Science, Department of Biomathematics and Computer Science, Istv\'an u. 2,
H-1078 Budapest, Hungary
\end{flushleft}
\pagebreak
\input psfig

\begin{abstract}

We compare the spectral properties
of long meteorological temperature records
with laboratory measurements in small convection cells.
Surprisingly, the atmospheric 
boundary layer sampled on a daily scale 
shares the statistical properties
of temperature fluctuations in small-scale experiments.
This fact can be explained by the hydrodynamical similarity 
between these seemingly very different systems.
The results suggest that the dynamics of daily
temperature fluctuations is determined by 
the soft turbulent state of the
atmospheric boundary layer in continental climate.

\end{abstract}

\bigskip

The studies of turbulence in the atmospheric boundary-layer
has a history of almost hundred years.
The theoretical description
of the basic mechanisms has been 
elaborated in detail \cite{1}, however the application of the 
concept of turbulence at the level of observed data evaluation is 
still not complete.
For example, one of the most common meteorological 
quantities measured
for a long time at various frequencies is the 
temperature. The length
and the relative accuracy of the available time series 
make these records
very attractive for detailed statistical analysis. 
Indeed,
there have been several attempts to describe the dynamics 
of daily medium
temperature fluctuations. A typical viewpoint is that 
the underlying mechanism
is fully stochastic in nature and can be considered as 
an autoregressive
process \cite{2}. A completely different viewpoint 
suggests that the apparent
irregularities may be attributed to a deterministic 
chaotic behavior, although
serious doubts have arisen on the existence of low 
dimensional chaos in the
long time behavior of the atmosphere (climatic 
attractor) \cite{3}, as well as 
in the processes over short time scales \cite{4}.

Our study based on daily temperature measurements of the Hungarian
Meteorological Service performed at twenty different 
meteorological stations covering the area of Hungary for the period 
1951 -- 1989. We compare the statistical
properties of our time series with measurements
on local temperature fluctuations in gaseous
helium and air \cite{6,7,8} performed by the Chicago Group.
A typical experimental setup for studying
free thermal convection consists of a closed
container filled
with different gases or liquids (Fig.~1), and the local
temperature inside is recorded on a millisecond scale.

Let us start with the characterization of the meteorological
data.
These records show a systematic seasonal 
periodicity and superimposed daily random fluctuations,
a detailed analysis has been published in Ref.~\cite{5}. We 
produced the fluctuation time series by subtracting 
the daily average over 39 years from the daily
medium temperature data. Thus,
the fluctuation $T_f(d,y)$ on day $d$ of 
year $y$ is given by the deviation of the actual daily
medium $T(d,y)$ from the seasonal average $\langle T\rangle (d)$, i.e.:
$$T_f(d,y)=T(d,y)-\langle T\rangle (d) \quad,$$
where $\langle T\rangle (d)$ is defined by
$$\langle T\rangle (d)=\frac{1}{39}\sum_{y=1951}^{1989}T(d,y) \quad .$$
The histogram of the fluctuation amplitudes has a
typical Gaussian distribution (Fig.~2). The power 
density spectrum of the time series measured at different 
meteorological stations can be fitted well by the
function (see Fig.~3):
$$ P(f) = P_s \exp \left[ - \left( {f\over 
f_s}\right)^\beta \right]\quad,
\eqno(1)$$
where $P_s=222\pm 16$, $f_s=0.017\pm 0.005$ day$^{-1}$, $\beta 
=0.54 \pm 0.03 $, here
the deviations indicate slight meteorological station 
dependence. Note that the end of the spectra could be
approximated with a power-law shape as well, but only
on a narrow frequency range (less than a half order
of magnitude). Surprisingly, the fitting form Eq.~(1) with
the same exponent $\beta$ and the Gaussian fluctuation
histogram are the same as was found in
experiments on convecting helium \cite{6,7},
in the so-called {\it soft turbulent} regime. How can one relate
atmospheric phenomena to small-scale model experiments?

It is well known that two hydrodynamical systems are 
similar, if their dimensionless
parameters are equal. In the case of experiments on thermal
convections, the analysis
is usually based upon the Boussinesq approximations\cite{1},
which give
two dimensionless parameters. First, the Rayleigh number
is
given by
$$ {\rm Ra} = {\alpha g L^3 \Delta T\over \kappa \nu} \quad, \eqno(2)
$$
where $\alpha$ is the (isobar) thermal expansion
coefficient,
$g$ is the gravitational acceleration, $L$ is the height
of the container, $\Delta T=T^+-T^-$ is the temperature
difference between
the top and bottom plates (Fig.~1), $\kappa$ is the thermal
diffusivity,
while $\nu$ is the kinematic viscosity. Second, the
Prandtl
number is simply the ratio
$$ {\rm Pr} = {\nu \over \kappa} \quad. \eqno(3)$$
These two numbers together with the aspect-ratio of the cell
($A=$width/height) give
a complete description of the experimental system.

The material parameters of the air do not depend
drastically on the
meteorological circumstances, some typical values are the
following \cite{1,10}:
$\alpha= 3.55\times 10^{-3}$ 1/K, $g=9.81$ m/s$^2$,
$\kappa = 2.03\times 10^{-5}$ m/s$^2$,
$\nu = 1.44\times 10^{-5}$ m/s$^2$. Between 0 and 40
$^\circ$C, the Prandtl
number remains constans: Pr$^{\rm air} = 0.71$.
The Prandtl
number of the helium gas
was kept also on a constant value Pr$^{\rm He}=0.64$ in the experiments
\cite{6,7}.

Three important points have to be considered for the 
proper comparison
of the observations, because the atmospheric boundary layer 
differs in some respects from the laboratory setups. 
First, there is no well-defined top plate above
the gas layer. 
This makes the term $L^3\Delta T$ in Eq.~(2) indefinite, therefore
the direct determination
of an atmospheric Rayleigh number is impossible. Fortunately, 
there are two characteristic frequencies in the power spectra, 
and a proper matching (see later) can provide an alternative
estimation for the length $L$. 
Second, the atmospheric boundary 
layer has practically an infinite aspect ratio, 
while the laboratory experiments
were performed in cells of low aspect ratios.
This problem can be 
resolved
by observations, which suggest that the vertical and 
horizontal sizes of the
medium scale ($\sim 100-500$ m) convective eddies are 
approximately
equal. Third, adequate measurements show that a 
characteristic turnover time of these eddies 
is in the range of minutes, while the sampling rate of the
meteorological records is one day.
Note, however, that the meteorological time series 
consist 
of daily {\it medium} temperature data, that is  
the mean of the daily
absolute maximum and minimum values.
Indeed, this can differ from the daily {\it average} temperature,
and may reflect of effects on much shorter time scale than one
day.

The control parameter in the laboratory experiments is the
Rayleigh number Ra. The Chicago Group covered the range
of Ra from $10^5$ to $10^{11}$ by changing the pressure of different
gases \cite{6,7,8}.
After the onset of convection, two different chaotic
domains Ra $=1.5-2.5\times 10^5$ and
Ra $=2.5-5\times 10^5$ with attractor dimensions
$D\approx 2$ and $D\approx 4$ were identified.
There are two regimes after the onset of turbulence:
The {\it soft turbulent} state ($5\times 10^5<$ Ra $<$ Ra$_c$)
and {\it hard turbulent} state (Ra $>$ Ra$_c$), where 
Ra$_c\approx 10^8$ depending slightly on the aspect ratio $A$.
The probability distribution of the local temperature
fluctuations
is Gaussian in soft turbulence, while exponential
in hard turbulence.
The power spectrum of the fluctuations is
stretched-exponential in soft turbulence [Eq.~(1)],
while it is
a power-law with an
exponent $-7/5$ in hard turbulence. The Gaussian amplitude
distribution (Fig.~2) and the stretched-exponential
power spectrum (Fig.~3) of the meteorological time series
suggest that the atmospheric boundary layer may exhibit 
soft turbulence, let us analyse further this assumption.

It is known from the experiments \cite{6,7,8}, how
the characteristic frequencies of the power spectra
depend on Ra. First, the fitted frequency $f_s$ of
Eq.~(1) (measured in units of $\kappa /L^2$) can
be estimated from Fig.~3 of Ref.~\cite{6} in the
soft turbulent regime as 
$ f_s({\rm Ra}) \approx 63\kappa /L^2 $ for Ra $\approx 10^6$,
and 
$ f_s({\rm Ra}) \approx 13000\kappa /L^2 $ for Ra $\approx 10^8$. 
Another characteristic value is the cutoff-frequency
$f_{max}$, at which the power spectra flatten out
as a consequence of noise (Fig.~4). 
This frequency depends also on Ra, moreover the 
relationship does not change at the transition of
soft--hard turbulence.
 From Fig.~13 of Ref.~\cite{8}, one can estimate
$ f_{max}({\rm Ra}) \approx 2500\kappa /L^2 $ for Ra $\approx 10^6$,
and
$ f_{max}({\rm Ra}) \approx 10^5\kappa /L^2 $ for Ra $\approx 10^8$.
The values for $f_{max}$  depend on the distance of thermistor from
the fix plate \cite{8}, but not to the extent that it could affect 
our following order of magnitude estimation.
Both the characteristic frequency $f_s=2\times 10^{-7}$
Hz of Eq.~(1) and
the the cutoff frequency $f_{max}=4.5\times 10^{-6}$ Hz of Fig.~4,
yield to an estimation $L\approx 80-1100$ m
in the atmospheric boundary layer. Note that this
seemingly wide interval belongs to a Rayleigh number range
of $10^6<$ Ra $10^8$. 
On the other hand, this height $L$ is in agreement with the
accepted values of
the thickness of the air layer influenced by the {\it daily
cycle} of temperature
change \cite{11}.

These results indicate that the typical atmospheric 
dynamics will unlikely
exhibit low dimensional chaos, since soft turbulent 
state occurs after 
several transitions from the chaotic state and is
connected with the increase 
of effective degrees of freedom. 
Furthermore, the present concept makes a strong 
prediction on daily temperature fluctuations at 
extreme climatical conditions. 
In such a case, the Rayleigh number can exceed the
threshold value Ra$_c$,
and then hard turbulence develops. 
Indeed, recent studies on North 
Pole data \cite{12} suggest, that characteristics 
of hard turbulence 
can be observed.

Finally, we would like to mention, that daily 
fluctuations of other meteorological data 
often show non-trivial scaling 
behaviour. For example, relative air 
humidity fluctuations have $1/f^\alpha$ ($\alpha\approx 
0.61$) power spectrum \cite{13},
for more than two decades of frequency range. Power-law 
scaling of frequency spectra 
is the typical sign of nontrivial complex behaviour and 
self-organization. This 
implicitly confirms the turbulence based 
explanation of fluctuations on a
daily scale, since density fluctuations, like those of 
air humidity, in turbulent 
flows, are expected to have power-law frequency scaling. 
It would be interesting
to analyze the daily fluctuations measured at different 
meteorological conditions, world
wide, from the point of view outlined here. By doing 
this, we could learn how different 
climates affect the transition between hard and soft 
turbulence and draw conclusions
how global warming, ice ages, natural catastrophes and 
other large scale climate changes 
affect the distribution of temperature measured on the 
ground. This information
could be useful, since a transition toward hard turbulence 
increases the occurrence of large 
daily fluctuations of order $\pm$ 20 $^\circ$C. A shift 
toward a power law frequency spectra 
increases the occurrence of extremely long correlated 
time periods when the temperature significantly 
deviates from the seasonal average values. 

This work has been supported by the "Nonlinear 
Phenomena" EU Research Network and the Hungarian
National Science Foundation (OTKA).



\begin{figure}
\centerline{\strut\psfig{figure=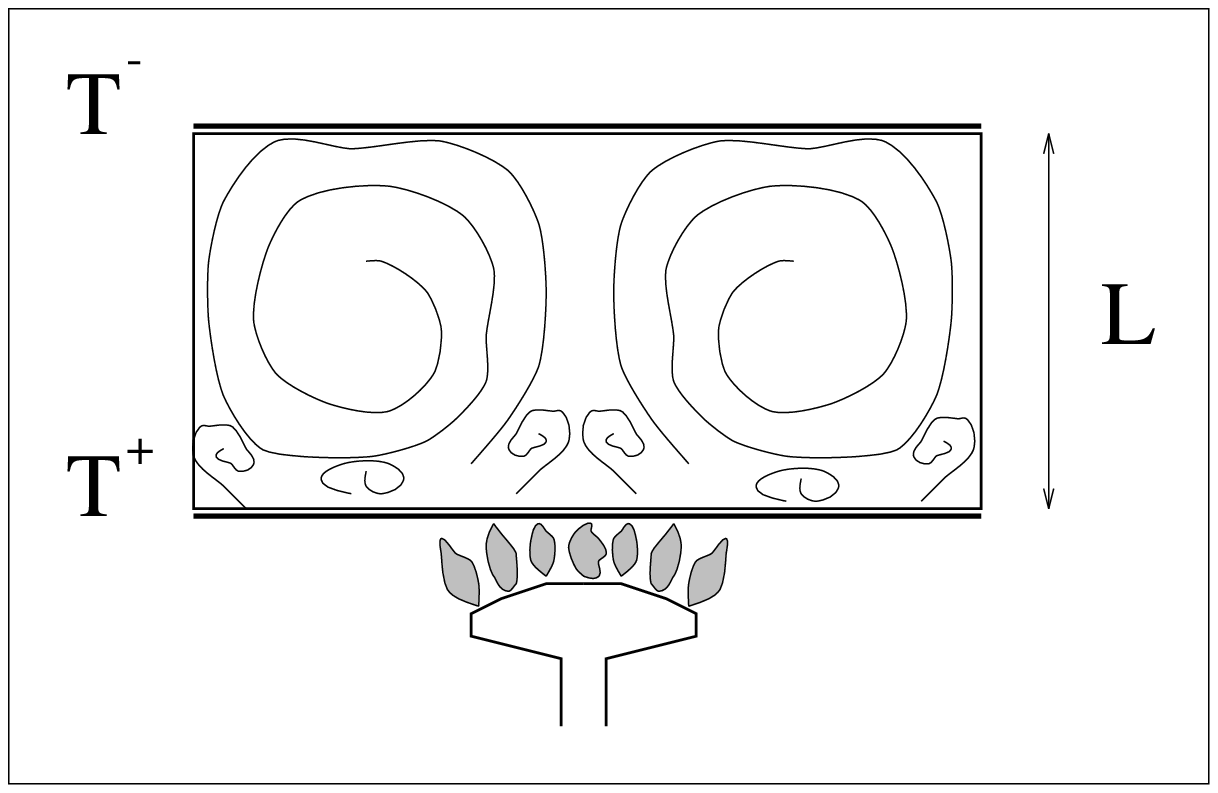,width=70mm}}
\caption{
Scematic of
typical experimental setups for studying
free thermal convection.
A characteristic size
of the gas container
is usually $L=10-40$ cm. The bottom plate is heated
($T^+$), while the
top plate is kept on a constants $T^-$ (lower)
temperature. The instantaneous gas temperature inside is measured 
by small thermistors placed to a desired location.
The atmospheric boundary layer bears a close resemblance
to this experiment: The air above the ground surface
is heated also from below and cooled at the top by higher
air layers.}
\end{figure}

\begin{figure}
\centerline{\strut\psfig{figure=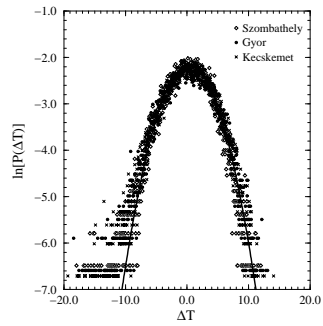,width=70mm}}
\caption{Normalized probability distributions for the detrended 
daily medium temperature fluctuations measured between 
1.1.1951--31.12.1989
in three different stations. The solid line shows a 
Gaussian fit with a
center of 0.31 $^\circ$C and standard deviation of 3.50 
$^\circ$C. 
The Gaussian shape together with the 
slight deviation at the left side (note the logarithmic 
scale) is the
same for all of the twenty evaluated records [5].}

\end{figure}

\begin{figure}
\centerline{\strut\psfig{figure=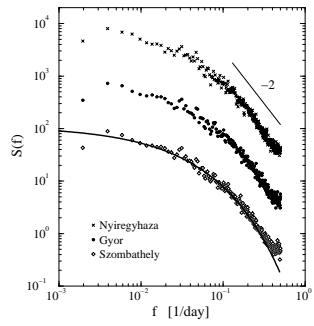,width=70mm}}
\caption{Unnormalized power density spectrum of the temperature 
fluctuations
measured in three stations.
 The frequency unit is day$^{-1}$. The upper two spectra are 
shifted upwards for the
sake of clearness. The result were obtained by a 
standard Fast Fourier Transformation 
method using Hann windowing [9].
The thick solid
line is the fit given by  Eq.~(1), the thin line 
illustrates an $1/f^2$
behavior on a restricted frequency range.}
\end{figure}

\begin{figure}
\centerline{\strut\psfig{figure=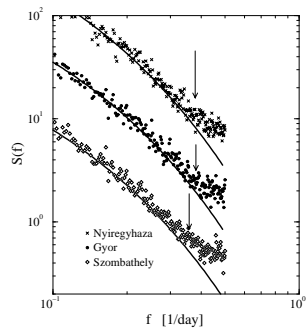,width=70mm}}
\caption{The same as Fig. 3 zoomed to the end of the spectra. 
The arrows indicate
the cutoff frequencies $f_{max}=0.39\pm 0.02$ day$^{-1}$, 
where the spectra reaches the
noise level.}
\end{figure}
\end{document}